\documentclass[10pt]{article}
\usepackage{graphicx}
\usepackage{amsmath}
\usepackage{amssymb}
\usepackage{caption2}
\begin{document}
\bibliographystyle{prsty}
\begin{center}
{\large {\bf \sc{ Analysis of the $B \to K^*_0(1430), \, a_0(1450)$ form-factors with light-cone QCD sum rules  }}} \\[2mm]
Zhi-Gang Wang \footnote{E-mail,wangzgyiti@yahoo.com.cn. }, Jun-Fang Li    \\
 Department of Physics, North China Electric Power University, Baoding 071003, P. R. China
\end{center}

\begin{abstract}
In this article, we take the scalar mesons $K^*_0(1430)$ and
$a_0(1450)$ as the conventional two-quark states,  and study the $B \to K^*_0(1430)
,  a_0(1450)$ form-factors with the $B$-meson light-cone QCD sum
rules,    then take those form-factors as  basic input  parameters to study
the semi-leptonic decays $\bar{B}^0\to a_0^+(1450)l \bar{\nu}_l $  and
 $B^-\to K_0^{*-}(1430)l\bar{l}$, the predictions can be confronted with  the experimental data in the future.
\end{abstract}



\section{Introduction}

There are many candidates  with the quantum numbers $J^{PC}=0^{++}$ below $2\,\rm{GeV}$, which
cannot be accommodated in one $\bar{q}q$ nonet, some are supposed to
be glueballs, molecular states and tetraquark states (or their
special superpositions)
\cite{Close2002,ReviewJaffe,ReviewAmsler2,KlemptPRT}.    The dynamics dominates  the $0^{++}$ mesons  below and above
$1\,\rm{GeV}$ respectively maybe different, and  there maybe exist  two scalar nonets below
$1.7\,\rm{ GeV}$ \cite{Close2002,ReviewJaffe,ReviewAmsler2}. The
strong attractions between the scalar diquark states $\mathbb{Q}^a$
and $\bar{\mathbb{Q}}^a$  in the relative $S$-wave maybe  result in a
nonet tetraquark states  manifest below $1\,\rm{GeV}$, while the
conventional $^3P_0$ $\bar{q}q$ nonet have masses about $(1.2-1.6)
\,\rm{GeV}$, and the well established  $^3P_1$ and $^3P_2$
$\bar{q}q$ nonets with $J^{PC}=1^{++}$ and $2^{++}$ respectively lie
in the same region. There are enough candidates for the
$^3P_0$ $\bar{q}q$ nonet mesons, $a_0(1450)$, $f_0(1370)$,
$K_0^*(1430)$, $f_0(1500)$ and $f_0(1710)$ \cite{PDG}. In
Ref.\cite{KlemptPRT}, Klempt and Zaitsev suggest that there maybe
exist four scalar nonets below $2.3\,\rm{GeV}$,
\begin{eqnarray}
\{f_0/\sigma(600),a_0(980),K^*_0(800),f_0(980) \}\, ,&&
\{f_0(xxx),a_0(1450),K^*_0(1430),f_0(1500) \}\, , \nonumber \\
\{f_0(xxx),a_0(xxx),K^*_0(xxx),f_0(1760) \}\, ,
&&\{f_0(xxx),a_0(2020),K^*_0(1950),f_0(2100) \} \, .
\end{eqnarray}
The lowest scalar  nonet  mesons $\{f_0/\sigma(600),a_0(980),K^*_0(800),f_0(980) \}$ are usually taken as the tetraquark
states, while the higher  scalar nonet mesons
$\{f_0(xxx),a_0(1450),K^*_0(1430),f_0(1500) \}$ are taken as the
conventional ${}^3P_0$ $\bar{q}q$ states \cite{Jaffe1977}. We should study such an scenario
 extensively before making definite conclusion,  it is interesting to study  the   scalar meson productions in the
  semi-leptonic $B$-decays.   Compared with the two-body hadronic $B$-decays, the semi-leptonic
$B$-decays suffer from much less theoretical uncertainties involving   the hadronic matrix elements.

 Experimentally, the semi-leptonic decays $\bar{B}^0\to a_0^+(1450)e \bar{\nu}_e $, $\bar{B}^0\to a_0^+(1450)\mu \bar{\nu}_{\mu}$,
$\bar{B}^0\to a_0^+(1450)\tau \bar{\nu}_{\tau} $, $B^-\to K_0^{*-}(1430)e^+ e^- $, $B^-\to K_0^{*-}(1430)\mu^+ \mu^-$,
$B^-\to K_0^{*-}(1430)\tau^+ \tau^- $,  which take place through the transitions $b\to ul\bar \nu_l$, $ sl\bar l$ at the quark level,  have not been observed yet \cite{PDG}. The theoretical predictions of the branching ratios based on  the QCD sum rules \cite{Aliev0710},  the perturbative QCD \cite{Lu0811},
 the light-cone QCD sum rules \cite{F-WangYM,F-Sun}  and
the covariant light-front quark model \cite{F-Chen} differ from each other greatly, another study using a different approach is worthy.

The processes induced by the flavor-changing neutral currents $b \to s(d)$
 are forbidden at the tree-level in the standard model, provide the most sensitive and
 stringent  test for the standard model at one-loop level, and can put
powerful constraints on the new physics models
\cite{BurasRMP,PRT2010,EPJC2008}. Experimentally, the semi-leptonic decays $B^0 \to \pi^- l^+\nu_l$, $ \rho^- l^+\nu_l$,
$B^+ \to \pi^0 l^+\nu_l$, $\eta l^+\nu_l$, $\eta' l^+\nu_l$, $ \rho^0 l^+\nu_l$, $ \omega l^+\nu_l$  have been observed \cite{PDG}.
In studying the semi-leptonic  $B$-decays to the pseudoscalar ($P$),
vector ($V$), axial-vector ($A$), tensor ($T$) and scalar ($S$) mesons,  we often encounter
the  $B \to P$, $V$, $T$, $S$ form-factors,
which are highly  nonperturbative quantities.
 The existing theoretical
works focus on the $B \to P,\,V$ form-factors, while the works on the $B\to T,\,S$ form-factors are relatively  few.
 It is worthy to study the $B \to S$ form-factors and
calculate the $B\to S\bar{l}l$, $Sl\bar{\nu}_l$ branching ratios to confront with the experimental data in the future.

In Refs.\cite{Khodjamirian05,KhodjamirianB07},  Khodjamirian et al
obtain new sum rules for the $B\to \pi$, $K$, $\rho$, $K^*$
form-factors from the correlation functions expanded near the
light-cone in terms of the $B$-meson distribution amplitudes, and
suggest QCD sum rules motivated models for the  three-particle
$B$-meson light-cone distribution amplitudes, which satisfy the
relations between the two-particle and three-particle $B$-meson
light-cone distribution amplitudes derived from the QCD equations of
motion and heavy-quark symmetry \cite{Qiao2001}. The new light-cone QCD sum
rules (or the $B$-meson light-cone QCD sum rules) have been applied  to calculate the $B\to
a_1(1260)$, $K^*_2(1430)$, $a_2(1320)$, $f_2(1270)$, $D$, $D^*$
form-factors \cite{WangPLB,KhodjamirianEPJC,Wang1011}.  Compared with the light-cone distribution amplitudes of
the light pseudoscalar mesons and
vector  mesons,  the $B$-meson light-cone distribution amplitudes
have  received relatively little attention. Our knowledge about the
nonperturbative parameters which determine the $B$-meson light-cone
distribution amplitudes is limited and an additional application
 based on the light-cone QCD sum rules  is useful.  In this article, we use the $B$-meson light-cone QCD sum rules to
calculate the $B \to K_0^*(1430)$, $a_0(1450)$ form-factors, and study the semi-leptonic $B$-decays.
We expect to extract the $B \to S$ form-factors from the experimental data on the semi-leptonic decays $B\to  S \bar{l} l$, $Sl\bar{\nu}_l$  at the
LHCb in the future, and obtain
severe constraints on the input parameters of the $B$-meson
light-cone distribution amplitudes.

The article is arranged as: in Sect.2, we derive the $B \to
K_0^*(1430)$, $a_0(1450)$  form-factors   with the $B$-meson light-cone QCD
sum rules; in Sect.3, we present the
 numerical results and discussions; and
Sect.4 is reserved for our conclusion.

\section{The $B \to K^*_0(1430), a_0(1450)$  form-factors  with light-cone QCD sum rules}

In the following, we write down the definitions for the $B \to
K^*_0(1430)$  form-factors  $A_1(q^2)$, $A_0(q^2)$, $T(q^2)$  and
$T_5(q^2)$ \cite{Aliev0710,Lu0811,Cheng0508},
\begin{eqnarray}
 \langle K^*_0(p)|\bar{s}(0)\gamma_\mu\gamma_5 b(0)|B(P)\rangle&=&-i\left\{
\left[(P+p)_\mu-\frac{m_B^2-m_{K^*_0}^2}{q^2}q_\mu\right]A_1(q^2)+\frac{m_B^2-m_{K^*_0}^2}{q^2}q_\mu A_0(q^2) \right\} \, ,\nonumber\\
 \langle K^*_0(p)|\bar{s}(0)\sigma_{\mu\nu}b(0)|B(P)\rangle&=&-i\epsilon_{\mu\nu\alpha\beta} (P+p)^\alpha q^\beta \frac{2T(q^2)}{m_B+m_{K^*_0}}
 \, ,\nonumber\\
 \langle K^*_0(p)|\bar{s}(0)\sigma_{\mu\nu}\gamma_5b(0)|B(P)\rangle&=&(q_\mu
p_\nu-p_\mu q_\nu)\frac{2T_5(q^2)}{m_B+m_{K^*_0}} \, ,
\end{eqnarray}
where $q_\mu=P_\mu-p_\mu$. In this article, we write down the calculations for the
$B\to K^*_0(1430)$ form-factors  explicitly, and obtain others via
the flavor $SU(3)$ symmetry for simplicity.

 We study the form-factors with the  two-point correlation functions $\Pi^k_{\mu}(p,q)$,
\begin{eqnarray}
\Pi^k_{\mu}(p,q)&=&i \int d^4x \, e^{i p \cdot x}
\langle 0 |T\left\{J(x) J^k_{\mu}(0)\right\}|B(P)\rangle \, ,\nonumber \\
J(x)&=&\bar{u}(x)s(x)\, ,\nonumber \\
J^1_\mu(x)&=&\bar{s}(x)\gamma_\mu\gamma_5 b(x)\, ,\nonumber \\
J^2_\mu(x)&=&\bar{s}(x)\sigma_{\mu\nu} b(x)z^\nu\, ,\nonumber \\
J^3_\mu(x)&=&\bar{s}(x)\sigma_{\mu\nu}\gamma_5 b(x)z^\nu \, ,
\end{eqnarray}
 where $k=1,2,3$, $P_\mu=p_\mu+q_\mu=m_Bv_\mu$, $v^2=1$, and $z_\mu$ is a four-vector.

According to the quark-hadron duality \cite{SVZ79,Reinders85}, we
can insert  a complete set of intermediate states with the same
quantum numbers as the current operator  $J(x)$ into the correlation
functions $\Pi^k_{\mu}(p,q) $ to obtain the hadronic representation.
After isolating the ground state contributions from the pole term of
the scalar meson $K_0^*(1430)$, we obtain the results:
\begin{eqnarray}
\Pi^1_{\mu}(p,q) &=&-\frac{if_{K^*_0}m_{K^*_0}}
  {m_{K^*_0}^2-p^2}
  \left[ \widetilde{A}_1(q^2)p_\mu+\widetilde{A}_2(q^2)P_\mu\right]   + \cdots \, , \nonumber \\
     \Pi^2_{\mu}(p,q) &=&-\frac{if_{K^*_0}m_{K^*_0}}
  {m_{K^*_0}^2-p^2}\frac{4T(q^2)}{m_B+m_{K^*_0}}
  \epsilon_{\mu\nu\alpha\beta}z^\nu p^\alpha P^\beta  + \cdots \, , \nonumber \\
  \Pi^3_{\mu}(p,q) &=&\frac{f_{K^*_0}m_{K^*_0}}
  {m_{K^*_0}^2-p^2}\frac{2T_5(q^2)}{m_B+m_{K^*_0}} \left(P_\mu p \cdot z-p_\mu P \cdot z\right)   + \cdots \, ,
    \end{eqnarray}
where
 \begin{eqnarray}
 \widetilde{A}_1(q^2)&=&\left(1+\frac{m_B^2-m_{K^*_0}^2}{q^2}\right)A_1(q^2)-\frac{m_B^2-m_{K^*_0}^2}{q^2}A_0(q^2) \, ,\nonumber\\
 \widetilde{A}_2(q^2)&=&\left(1-\frac{m_B^2-m_{K^*_0}^2}{q^2}\right)A_1(q^2)+\frac{m_B^2-m_{K^*_0}^2}{q^2}A_0(q^2)\, ,
 \end{eqnarray}
 and  the  decay constant $f_{K^*_0}$ is defined by $ \langle0|J(0)|K_0^*(p)\rangle = f_{K^*_0}m_{K^*_0} $.
   In this article, we derive the QCD sum rules with the tensor structures
$P_\mu$, $p_\mu$, $\epsilon_{\mu\nu\alpha\beta}z^\nu p^\alpha
P^\beta$  and $P_\mu p \cdot z-p_\mu P \cdot z$, respectively.

 In the following, we briefly outline
the operator product expansion for the correlation functions
$\Pi^k_{\mu}(p,q)$ in perturbative QCD. The calculations are
performed at the large space-like momentum region $p^2\ll 0$ with the constraint
$0\leq q^2< m_b^2+m_b p^2/\bar{\Lambda}$, where
$m_B=m_b+\bar{\Lambda}$ in the heavy quark limit. We write down the
propagator of a massive  quark in the external gluon field in the
Fock-Schwinger gauge and the $B$-meson light-cone distribution
amplitudes firstly \cite{KhodjamirianB07,Belyaev94},
\begin{eqnarray}
S_{ij}(x,y)&=&
 i \int\frac{d^4k}{(2\pi)^4}e^{-ik(x-y)}\left\{
\frac{\not\!k +m}{k^2-m^2} \delta_{ij} -\int\limits_0^1 du\, g_s \,
G^{\mu\nu}_{ij}(ux+(1-u)y)\right.\nonumber\\
 &&\left. \left[ \frac12 \frac {\not\!k
+m}{(k^2-m^2)^2}\sigma_{\mu\nu} - \frac1{k^2-m^2}u(x-y)_\mu
\gamma_\nu \right]\right\}\, ,
\end{eqnarray}

\begin{eqnarray}
 \langle 0|\bar{q}_{\alpha}(x)
h_{v\beta}(0) |B(v)\rangle &=& -\frac{if_B m_B}{4}\int\limits
_0^\infty d\omega e^{-i\omega v\cdot x}  \left \{(1 +\not\!v) \left
[ \phi_+(\omega) - \frac{\phi_+(\omega) -\phi_-(\omega)}{2 v\cdot
x}\not\! x \right ]\gamma_5\right\}_{\beta\alpha} \, , \nonumber
\end{eqnarray}

\begin{eqnarray}
\langle 0|\bar{q}_\alpha(x) G_{\lambda\rho}(ux)
h_{v\beta}(0)|B(v)\rangle &=& \frac{f_Bm_B}{4}\int\limits_0^\infty
d\omega \int\limits_0^\infty d\xi  e^{-i(\omega+u\xi) v\cdot
x}\left\{(1 + \not\!v) \Big[
(v_\lambda\gamma_\rho-v_\rho\gamma_\lambda)
 \right.
\nonumber \\
&&  \left(\Psi_A(\omega,\xi)-\Psi_V(\omega,\xi)\right)
-i\sigma_{\lambda\rho}\Psi_V(\omega,\xi)-\frac{x_\lambda
v_\rho-x_\rho v_\lambda}{v\cdot x}X_A(\omega,\xi)\nonumber\\
&&\left. +\frac{x_\lambda \gamma_\rho-x_\rho \gamma_\lambda}{v\cdot
x}Y_A(\omega,\xi)\Big]\gamma_5\right\}_{\beta\alpha}\,,
\end{eqnarray}
where
\begin{eqnarray}
\phi_+(\omega)&=&
\frac{\omega}{\omega_0^2}e^{-\frac{\omega}{\omega_0}} \, ,\,\,\,
\phi_-(\omega)= \frac{1}{\omega_0}e^{-\frac{\omega}{\omega_0}} \, ,\nonumber\\
 \Psi_A(\omega,\xi)& =& \Psi_V(\omega,\xi) = \frac{\lambda_E^2
}{6\omega_0^4}\xi^2 e^{-\frac{\omega+\xi}{\omega_0}} \, ,\nonumber\\
X_A(\omega,\xi)& = & \frac{\lambda_E^2 }{6\omega_0^4}\xi(2\omega-\xi)e^{-\frac{\omega+\xi}{\omega_0}}\,,\nonumber\\
Y_A(\omega,\xi)& =&  -\dfrac{\lambda_E^2 }{24\omega_0^4}
\xi(7\omega_0-13\omega+3\xi)e^{-\frac{\omega+\xi}{\omega_0}}\,,
\end{eqnarray}
the $\omega_0$ and $\lambda^2_E$ are some parameters of the
$B$-meson light-cone distribution amplitudes, then
 contract the $s$-quark field in the correlation functions
$\Pi^k_{\mu}(p,q)$ with Wick theorem to obtain the $s$-quark propagator, extract  the $B$-meson light-cone
distribution amplitudes, and carry out   the integrals over the variables
$x$ and $k$, finally  obtain the spectral densities  at the level of
quark-gluon degrees of freedom.  In this article, we take the
three-particle  $B$-meson light-cone distribution amplitudes
suggested in Ref.\cite{KhodjamirianB07}, they obey the powerful
constraints derived in Ref.\cite{Qiao2001} and the relations between
the matrix elements of the local operators and the moments of the
light-cone distribution amplitudes.
In the region of small $\omega$, the exponential form of
distribution amplitude $\phi_+(\omega)$ is numerically close to the
more elaborated model suggested in Ref.\cite{Braun2004},
\begin{eqnarray}
\phi_+(\omega, \mu=1 \mbox{GeV}) =\frac{4\omega}{\pi\lambda_B(1+\omega^2)} \left(\frac{1}{1+\omega^2}-2
\frac{\sigma_B-1}{\pi^2} \ln\omega\right)\, ,
 \end{eqnarray}
 where $\omega_0=\lambda_B$,   the $\omega$ is in unit of $\rm{GeV}$, and the $\sigma_B$ is a parameter.

We match the spectral densities at the hadron-level and quark-level below the continuum thresholds $s_0$, perform the Borel transformation with respect
to the variable $-p^2$, and obtain the QCD sum rules for the $B \to K^*_0(1430)$ form-factors
$\widetilde{A}_{+}(q^2)$, $\widetilde{A}_{-}(q^2)$, $T(q^2)$ and
$T_5(q^2)$, respectively,
\begin{eqnarray}
\widetilde{A}_\pm(q^2) &=& \frac{f_B m_B^2}{f_{K^*_0}
m_{K^*_0}}e^{\frac{m_{K^*_0}^2}{M^2}}\left\{
\int_{0}^{\sigma_0}d\sigma
\left\{\frac{\phi_+(\omega')}{\bar{\sigma}}\left(1\mp\frac{\omega'+m_s}{m_B}\right)+\left[\widetilde{\phi}_+(\omega')-\widetilde{\phi}_-(\omega')\right]
\right.\right. \nonumber \\
&& \left.\left[ \frac{m_s}{\bar{\sigma}^2M^2}\left(1\mp
\frac{\omega'}{m_B}\right)\mp\frac{1}{\bar{\sigma}m_B}\mp\frac{m_s^2}{\bar{\sigma}^2M^2m_B}\right]
\right\}e^{-\frac{s}{M^2}}
+\int_{0}^{\sigma_0}\widetilde{d\sigma}\left\{
\frac{\Psi_A(\omega,\xi)-\Psi_V(\omega,\xi)}{\bar{\sigma}^2}\right.\nonumber\\
&&\left[ \frac{1+2u}{M^2}\pm
\frac{3[(1-2u)\widetilde{\omega}-m_s]}{m_BM^2}\pm\frac{2(1-u)}{m_B^2}\left(1-\frac{\widetilde{m}_B^2}{M^2}\right)\right]
 +\frac{6\Psi_V(\omega,\xi)}{\bar{\sigma}^2M^2}\left(u\mp
\frac{m_s+u\widetilde{\omega}}{m_B} \right)  \nonumber \\
&&+\frac{2uX(\omega,\xi)}{\bar{\sigma}^2M^2}\left(1\mp\frac{\widetilde{\omega}}{m_B}
\right)+\frac{6m_s
\widetilde{Y}(\omega,\xi)}{\bar{\sigma}^3M^4}\left[-1
\pm\frac{\widetilde{\omega}+(2u-1)m_s}{m_B} \right]
+\frac{\widetilde{X}_A(\omega,\xi)}{\bar{\sigma}^2M^2} \left[
\pm\frac{3}{m_B} \right.\nonumber \\
&&\left. \left.+\frac{2(\widetilde{\omega}+m_s)}{\bar{\sigma}M^2}
\mp\frac{4}{m_B\bar{\sigma}}\left( 1-\frac{s}{2M^2}\right)
+\frac{2}{m_B\bar{\sigma}}\left(
1-\frac{\widetilde{m}_B^2}{2M^2}\right)\left(1\pm
\frac{\widetilde{\omega}+m_s}{m_B}\right)
 \mp \frac{4um_s^2}{\bar{\sigma}M^2m_B}\right]
\right\}  \nonumber \\
&&\left.e^{-\frac{s}{M^2}}\right\} \, ,
\end{eqnarray}
\begin{eqnarray}
T(q^2) &=& \frac{f_Bm_B(m_B+m_{K^*_0})}{4f_{K^*_0}
m_{K^*_0}}e^{\frac{m_{K^*_0}^2}{M^2}}\left\{
\int_{0}^{\sigma_0}d\sigma
\left\{\frac{\phi_+(\omega')}{\bar{\sigma}}+\frac{m_s}{\bar{\sigma}^2M^2}\left[\widetilde{\phi}_+(\omega')-\widetilde{\phi}_-(\omega')\right]
\right\}e^{-\frac{s}{M^2}}\right. \nonumber \\
&&+\int_{0}^{\sigma_0}\widetilde{d\sigma}\left\{
\frac{1+2u}{\bar{\sigma}^2M^2}\left[\Psi_A(\omega,\xi)-\Psi_V(\omega,\xi)\right]\right.
 +\frac{6u\Psi_V(\omega,\xi)}{\bar{\sigma}^2M^2} +\frac{2uX(\omega,\xi)}{\bar{\sigma}^2M^2}  \nonumber \\
&&\left.\left.-\frac{6m_s
\widetilde{Y}(\omega,\xi)}{\bar{\sigma}^3M^4}
+\frac{2\widetilde{X}_A(\omega,\xi)}{\bar{\sigma}^3M^2} \left[
 \frac{\widetilde{\omega}+m_s}{M^2}
+\frac{1}{m_B}\left( 1-\frac{\widetilde{m}_B^2}{2M^2}\right) \right]
\right\}  e^{-\frac{s}{M^2}}\right\} \, ,
\end{eqnarray}
\begin{eqnarray}
T_5(q^2)&=&2T(q^2) \, ,
\end{eqnarray}
where
\begin{eqnarray}
\widetilde{A}_\pm(q^2)&=&\widetilde{A}_1(q^2)\pm \widetilde{A}_2(q^2)\, ,\nonumber\\
\int_{0}^{\sigma_0}\widetilde{d\sigma}&=&\int_{0}^{\sigma_0}d\sigma
\int_0^{\sigma m_B}d\omega \int_{\sigma m_B-\omega}^\infty
\frac{d\xi}{\xi} \, ,\nonumber\\
s &=& m_B^2 \sigma-\frac{\sigma q^2-m_s^2}{\bar{\sigma}}\, , \,\,\,\omega'=\sigma m_B \, , \, \, \,\bar{\sigma}=1-\sigma \, , \, \, \, \widetilde{\omega} =\omega+u\xi\, ,\nonumber\\
\sigma_0&=&\frac{s^{K^*_0}_0+m_B^2-q^2-\sqrt{(s^{K^*_0}_0+m_B^2-q^2)^2-4(s^{K^*_0}_0-m_s^2)m_B^2}}{2m_B^2} \, ,\nonumber\\
u&=&\frac{\sigma m_B-\omega}{\xi} \, ,\,\,\,\widetilde{m}_B^2=m_B^2(1+\sigma)-\frac{q^2-m_s^2}{\bar{\sigma}}\, ,\nonumber\\
\widetilde{X}_A(\omega,\xi)&=&\int_0^\omega d\lambda
X_A(\lambda,\xi) \, ,\,\,\,\widetilde{Y}_A(\omega,\xi)=\int_0^\omega
d\lambda
Y_A(\lambda,\xi) \, ,\nonumber\\
\widetilde{\phi}_\pm(\omega)&=&\int_0^\omega d\lambda
\phi_\pm(\lambda) \, .
\end{eqnarray}
With a simple replacement, $ m_s\to 0$, $m_{K^*_0}\to m_{a_0}$, $f_{K^*_0}\to
 f_{a_0}$, $s_0^{K^*_0} \to s_0^{a_0}$, we can obtain the corresponding QCD sum rules for the $B\to a_0(1450)$
 form-factors.

\section{Numerical results and discussions}
The input parameters for the $B$-meson light-cone distribution
amplitudes are taken as
$\omega_0=\lambda_B(\mu)=(0.46\pm0.11)\,\rm{GeV}$, $\mu=1\,\rm{GeV}$
\cite{Braun2004}, $\lambda_E^2=(0.11\pm0.06)\,\rm{GeV}^2$
\cite{Grozin1997}, $m_B=5.279 \,\rm{GeV}$, and  $f_B=(0.18 \pm
0.02)\,\rm{GeV}$ \cite{LCSRreview}. The dominating contributions in
the QCD sum rules come from the two-particle $B$-meson light-cone
distribution amplitudes, the contributions from the three-particle
$B$-meson light-cone distribution amplitudes are of minor importance
as they are suppressed by additional powers of $\frac{1}{M^2}$. The
main uncertainty comes from the parameter $\omega_0$ (or $\lambda_B$),
which determines the line shapes of the two-particle and
three-particle light-cone distribution amplitudes of the $B$-meson.
In this article, we take the value from the QCD sum rules
 \cite{Braun2004}, where the $B$-meson light-cone distribution
amplitude $\phi_+(\omega)$ is parameterized  by the matrix element of the
bilocal operator at imaginary light-cone separation.

The masses, decay constants, threshold parameters and Borel
parameters for the scalar mesons $K^*_0(1430)$ and  $a_0(1450)$ are
determined by the conventional two-point QCD sum rules with the
scalar currents $\bar{u}(x)s(x)$ and $\bar{u}(x)d(x)$, respectively.
The values are  $m_{K^*_0}=(1.435\pm0.065)\,\rm{GeV}$,
$m_{a_0}=(1.470\pm0.070)\,\rm{GeV}$,
$f_{K^*_0}=(0.435\pm0.015)\,\rm{GeV}$,
$f_{a_0}=(0.460\pm0.015)\,\rm{GeV}$,
 $s_0^{K^*_0}=(4.0\pm0.2)\,\rm{GeV}^2$,
$s_0^{a_0}=(4.4\pm0.2)\,\rm{GeV}^2$, and
$M^2=(1.2-1.8)\,\rm{GeV}^2$, $(1.3-1.9)\,\rm{GeV}^2$
 in the channels $K^*_0(1430)$, $a_0(1450)$,
respectively.  In Ref.\cite{YangK1430}, Du, Li and Yang
 perform detailed studies of
 the mass and decay constant of the isospin $I=1/2$ scalar mesons composed of
 $s\bar{q}$ or $q\bar{s}$  using the QCD sum rules, and observe that the
  $K_{0}^{\ast}(1430)$ is the lowest scalar state  and
  the first radial excited state has the mass larger than $2.0\,\rm{GeV}$.
Close and Tornqvist  propose that  the lowest scalar nonet mesons
$\{f_0/\sigma(600),a_0(980),K^*_0(800),f_0(980) \}$  are tetraquark
states  consist of the scalar diquarks in the relative $S$-wave near the center, with some $\bar{q}q$ components in the relative
 $P$-wave, but further out they rearrange to form two colorless
$ \bar{q}q$ pairs and finally as the meson-meson states \cite{Close2002}.
The contaminations from the  scalar  mesons $K^*_0(800)$ and
$a_0(980)$ are very small if there are some.

Taking into account all uncertainties of the relevant parameters, we
obtain the numerical values of the  form-factors
 $A_1(q^2)$, $A_0(q^2)$,
 $T(q^2)$ and $T_5(q^2)$, which are shown in Fig.1 at zero momentum
 transfer. From the figure, we can see that the values of the form-factors
 are very stable with variations of the Borel
parameter.  The form-factors can be parameterized  in the double-pole form,
 \begin{eqnarray}
 F_i(q^2)&=&\frac{F_i(0)}{1+a_Fq^2/m_B^2+b_Fq^4/m_B^4}  \, ,
 \end{eqnarray}
where the $F_i(q^2)$ denote the $A_1(q^2)$, $A_0(q^2)$, $T(q^2)$ and
$T_5(q^2)$,  the $a_F$ and $b_F$ are the corresponding coefficients
and their values are presented in Table 1.

In Table 2, we also present the values of the form-factors from
other theoretical calculations, such as
 the QCD sum rules \cite{Aliev0710}, the perturbative QCD \cite{Lu0811},  the light-cone QCD  sum rules
\cite{F-WangYM,F-Sun},   and the covariant light-front quark model
 \cite{F-Chen}, the central values differ widely. If we take into account the large uncertainties,  there are some  overlaps among the predictions
   from different theoretical approaches. In this article, we will focus on the central values.

\begin{table}
\begin{center}
\begin{tabular}{|c|c|c|}\hline\hline
                  & $F_{BK^*_0}(0)$ $[a_F,b_F]$             & $F_{Ba_0}(0)$ $[a_F,b_F]$  \\ \hline
          $A_1$   & $0.42^{+0.25}_{-0.13}$ $[-1.50,0.42]$   & $0.45^{+0.25}_{-0.13}$ $[-1.46,0.39]$ \\ \hline
          $A_0$   & $0.42^{+0.25}_{-0.13}$ $[-0.42,0.19]$   & $0.45^{+0.25}_{-0.13}$ $[-0.43,0.16]$\\ \hline
          $T$     & $0.27^{+0.16}_{-0.09}$ $[-1.55,0.44]$   & $0.28^{+0.16}_{-0.08}$ $[-1.52,0.40]$\\ \hline
          $T_5$   & $0.54^{+0.32}_{-0.18}$ $[-1.55,0.44]$   & $0.56^{+0.32}_{-0.16}$ $[-1.52,0.40]$\\ \hline
  \hline
\end{tabular}
\end{center}
\caption{ The values of the form-factors at zero  momentum transfer
and the parameters of the fitted form-factors.}
\end{table}

\begin{table}
\begin{center}
\begin{tabular}{|c|c|c|c|c|c|c|}\hline\hline
                       &Ref.\cite{Aliev0710}  &Ref.\cite{Lu0811}       &Ref.\cite{F-WangYM}     &Ref.\cite{F-Sun} &Ref.\cite{F-Chen}  &This Work\\ \hline
    $A^{BK_0^*}_1(0)$  &$0.31 \pm 0.08$       &$0.60^{+0.18}_{-0.15}$  &$0.49^{+0.10}_{-0.10}$  &$0.49$           &$0.26$             &$0.42^{+0.25}_{-0.13}$ \\ \hline
    $T^{BK_0^*}_5(0)$  &$ 0.26 \pm 0.07$      &$0.78^{+0.25}_{-0.19}$  &$0.60^{+0.14}_{-0.13}$  &$0.69$           &$0.34$             &$0.54^{+0.32}_{-0.18}$ \\ \hline
     $A^{Ba_0}_1(0)$   &                      &$0.68^{+0.19}_{-0.15}$  &$0.52^{+0.10}_{-0.10}$  &                 &                   &$0.45^{+0.25}_{-0.13}$ \\ \hline
      $T^{Ba_0}_5(0)$  &                      &$0.92^{+0.30}_{-0.21}$  &$0.66^{+0.13}_{-0.14}$  &                 &                   &$0.56^{+0.32}_{-0.16}$ \\ \hline
   \hline
\end{tabular}
\end{center}
\caption{ The values of the form-factors at zero momentum transfer
from different theoretical approaches.}
\end{table}

\begin{table}
\begin{center}
\begin{tabular}{|c|c|c|c|c|c|}\hline\hline
                                                 &Ref.\cite{Aliev0710} &Ref.\cite{Lu0811}   &Ref.\cite{F-WangYM} &Ref.\cite{F-Chen}   & This Work \\ \hline
  $\bar{B}^0\to a_0^+(1450)e \bar{\nu}_e$        &                     &$3.25\times10^{-4}$ &$1.8\times10^{-4}$  &                    &$1.59\times 10^{-4}$ \\ \hline
  $\bar{B}^0\to a_0^+(1450)\mu \bar{\nu}_{\mu}$  &                     &$3.25\times10^{-4}$ &$1.8\times10^{-4}$  &                    &$1.59\times 10^{-4}$ \\ \hline
  $\bar{B}^0\to a_0^+(1450)\tau \bar{\nu}_{\tau}$&                     &$1.32\times10^{-4}$ &$6.3\times10^{-5}$  &                    &$5.83\times 10^{-5}$ \\  \hline
  $B^-\to K_0^{*-}(1430)e^+ e^- $                &$2.09\times10^{-7}$  &$9.78\times10^{-7}$ &$5.7\times10^{-7}$  &$1.63\times10^{-7}$ &$4.51\times 10^{-7}$ \\ \hline
  $B^-\to K_0^{*-}(1430)\mu^+ \mu^-$             &$2.07\times10^{-7}$  &$9.78\times10^{-7}$ &$5.6\times10^{-7}$  &$1.62\times10^{-7}$ &$4.48\times 10^{-7}$ \\ \hline
  $B^-\to K_0^{*-}(1430)\tau^+ \tau^-$           &$1.70\times10^{-9}$  &$6.29\times10^{-9}$ &$9.8\times10^{-9}$  &$2.86\times10^{-9}$ &$7.35\times 10^{-9}$ \\ \hline
   \hline
\end{tabular}
\end{center}
\caption{ The branching ratios from different theoretical approaches.}
\end{table}

\begin{figure}
 \centering
 \includegraphics[totalheight=5cm,width=6cm]{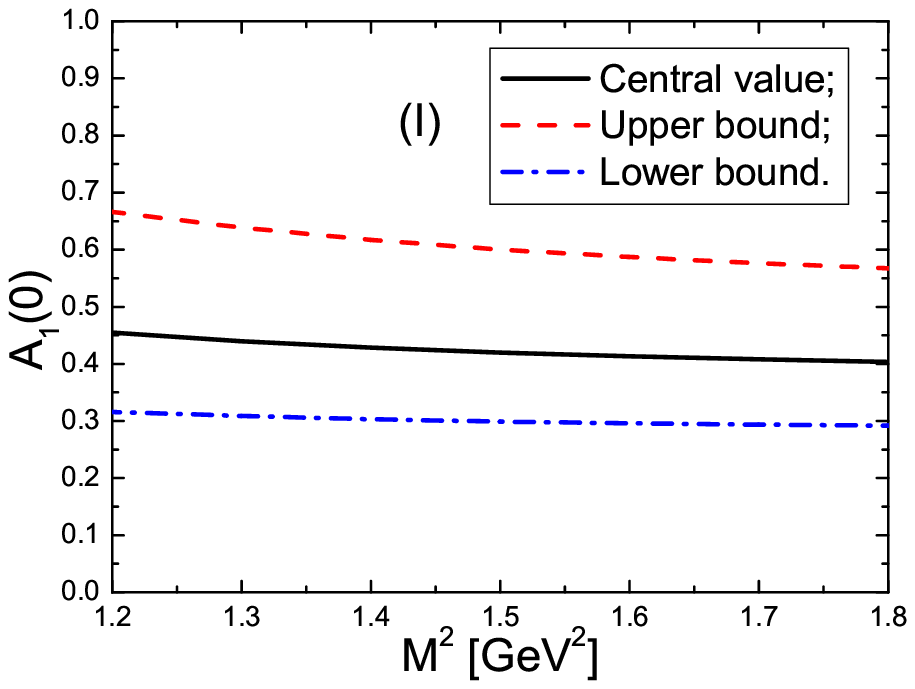}
 \includegraphics[totalheight=5cm,width=6cm]{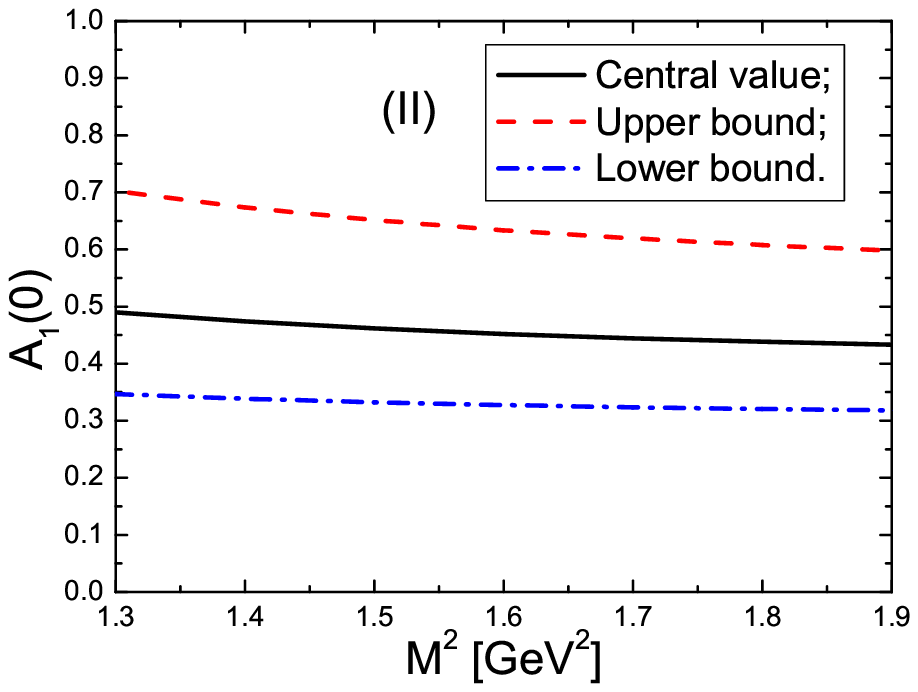}
 \includegraphics[totalheight=5cm,width=6cm]{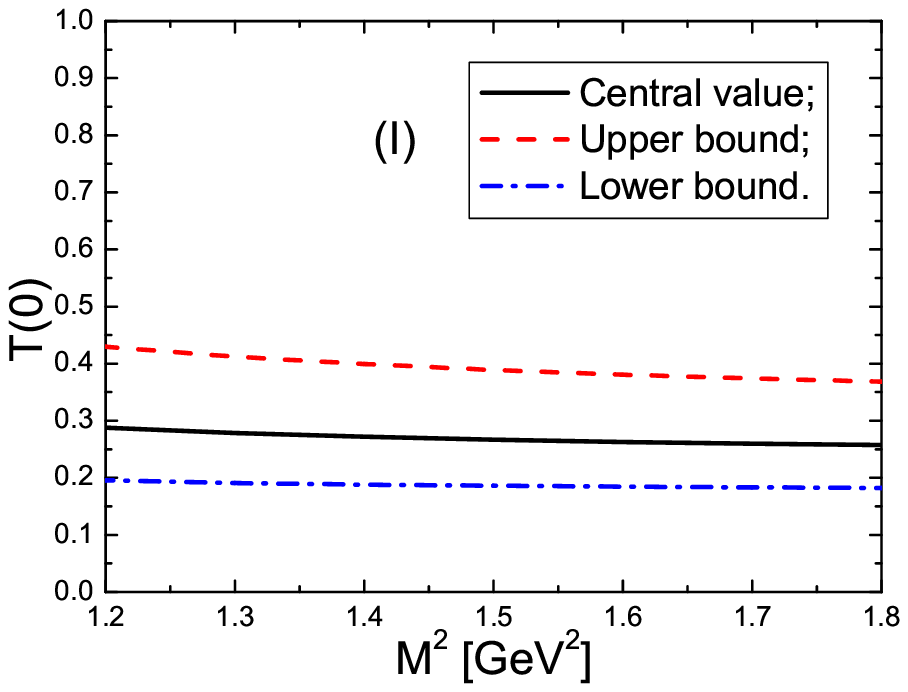}
  \includegraphics[totalheight=5cm,width=6cm]{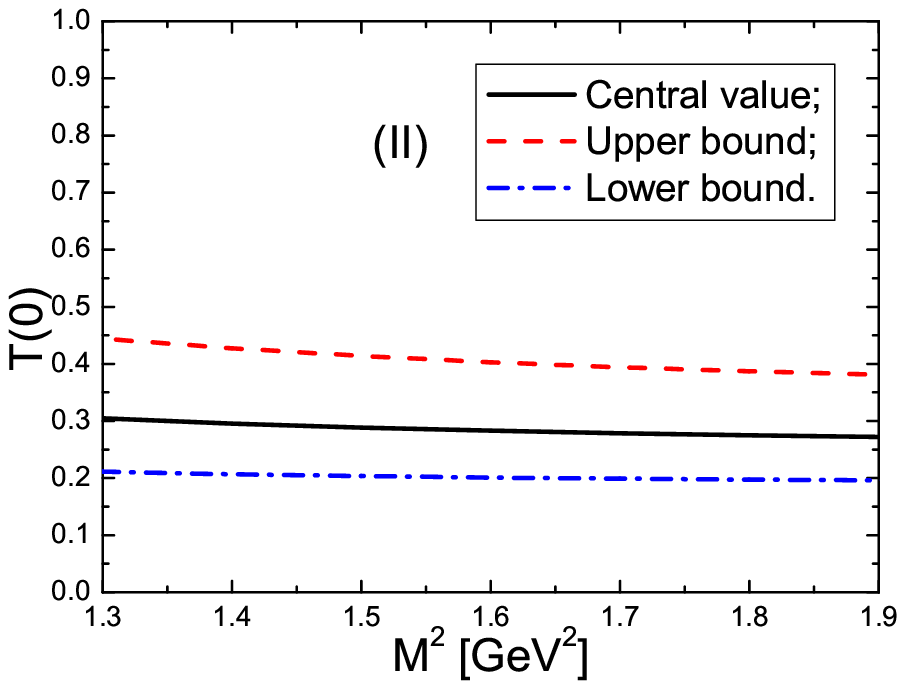}
    \caption{ The values of the form-factors at zero  momentum transfer
 with variation of the Borel  parameter $M^2$, the (I) and (II) denote the
 transitions  $B\to K^*_0(1430)$ and $B\to a_0(1450)$,  respectively. $A_0(0)=A_1(0)$ and $T_5(0)=2T(0)$.   }
\end{figure}

\begin{figure}
 \centering
 \includegraphics[totalheight=5cm,width=6cm]{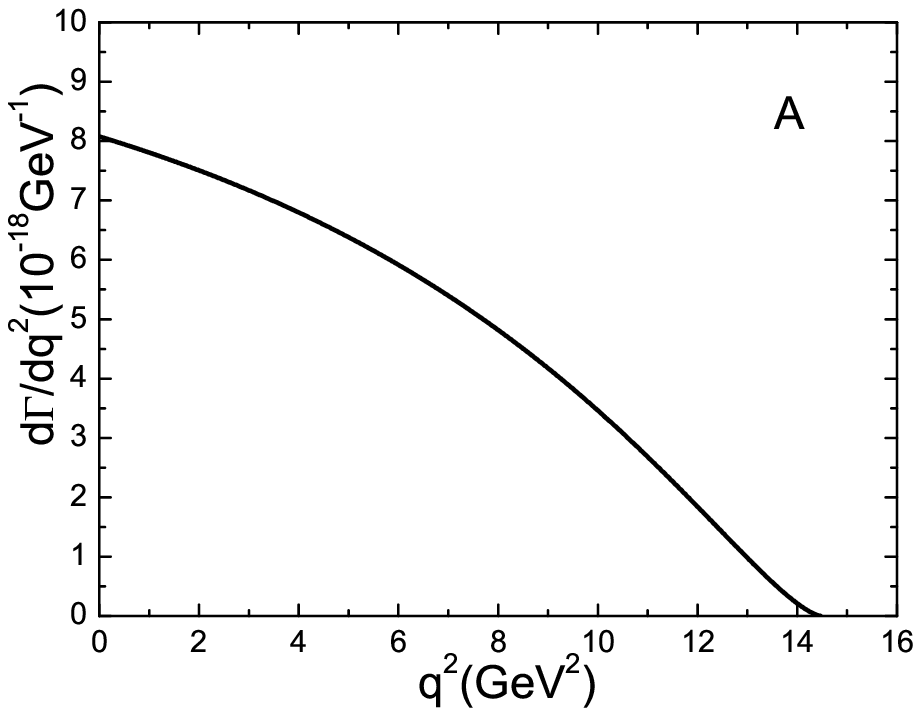}
\includegraphics[totalheight=5cm,width=6cm]{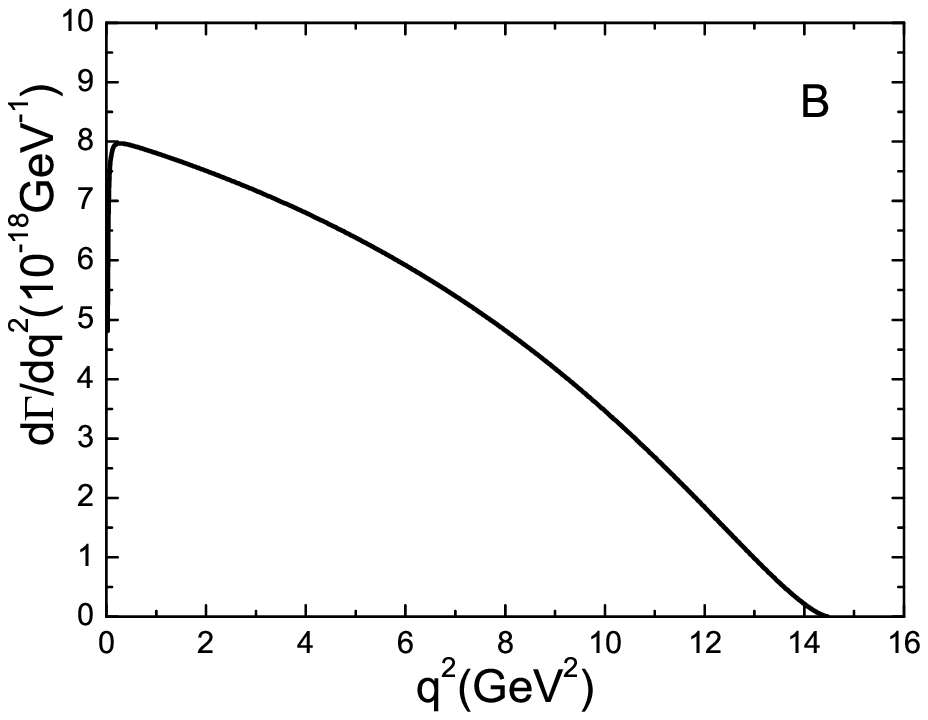}
\includegraphics[totalheight=5cm,width=6cm]{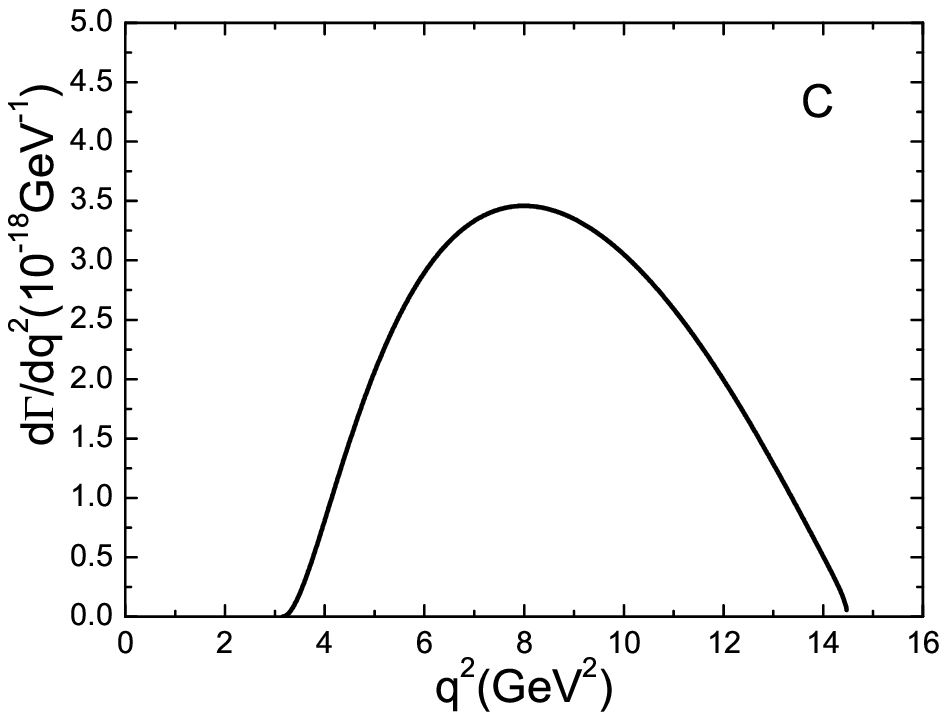}
\includegraphics[totalheight=5cm,width=6cm]{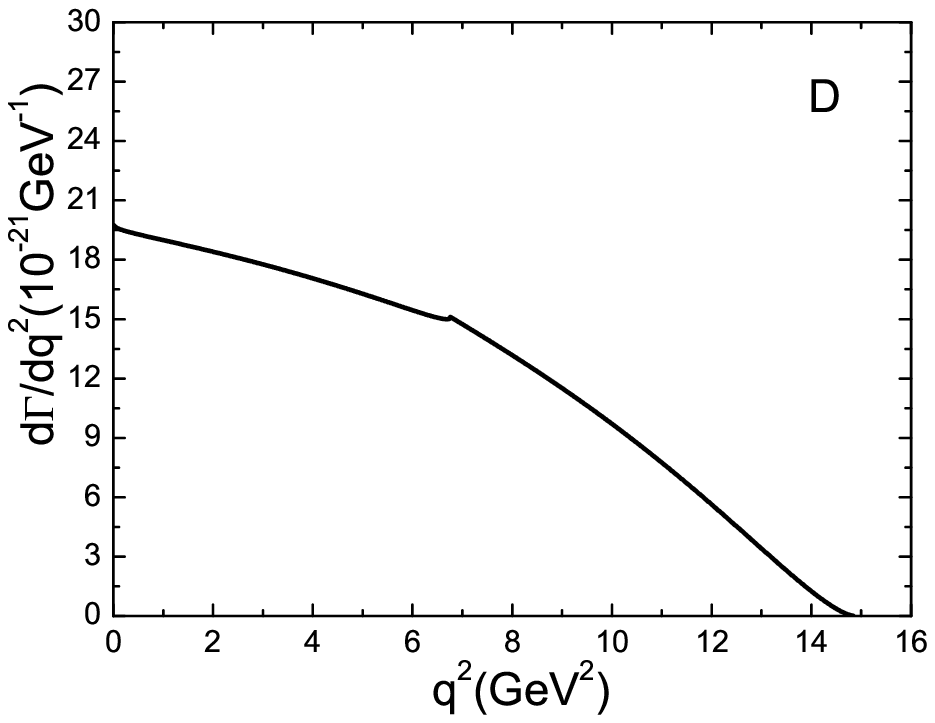}
\includegraphics[totalheight=5cm,width=6cm]{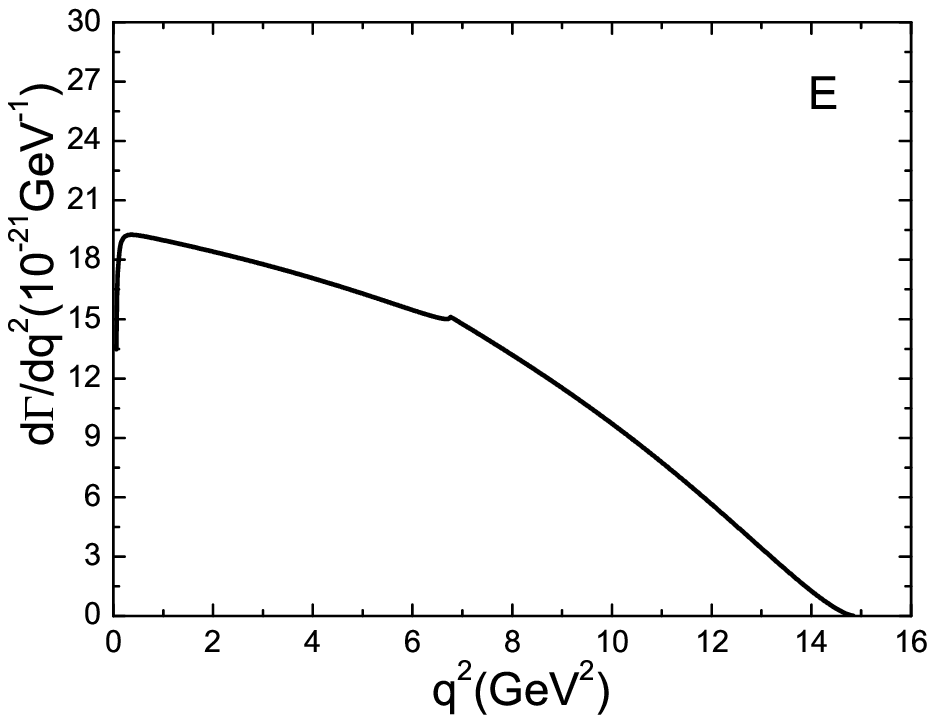}
\includegraphics[totalheight=5cm,width=6cm]{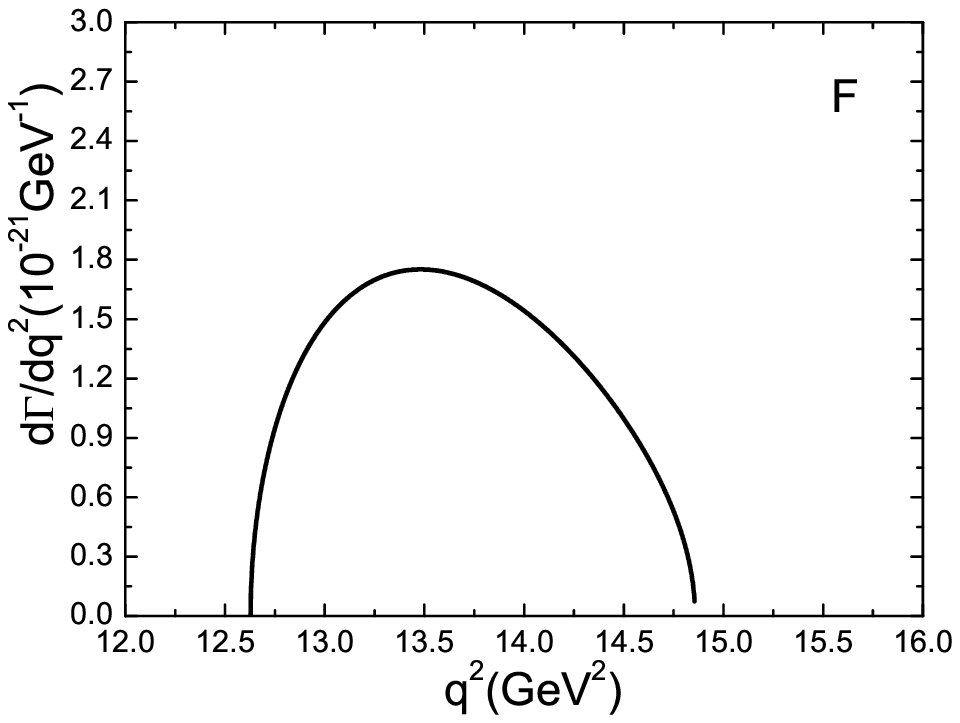}
    \caption{ The  values  of the differential decay widthes, the $A$, $B$, $C$, $D$, $E$ and $F$
 denote the decays   $\bar{B}^0\to a_0^+(1450)e \bar{\nu}_e$, $\bar{B}^0\to a_0^+(1450)\mu \bar{\nu}_{\mu}$,
 $\bar{B}^0\to a_0^+(1450)\tau \bar{\nu}_{\tau}$, $B^-\to K_0^{*-}(1430)e^+ e^- $, $B^-\to K_0^{*-}(1430)\mu^+ \mu^-$
  and $B^-\to K_0^{*-}(1430)\tau^+ \tau^-$,  respectively.   }
\end{figure}

We can take those form-factors as the basic input parameters and study the semi-leptonic decays to the scalar mesons $K^*_0(1430)$ and $a_0(1450)$. The effective Hamiltonians for the $b\to ul\bar \nu_l$ and $b\to sl\bar l$ transitions  can be written as
 \begin{eqnarray}
 {\cal H}_{eff}(b\to u l {\bar{\nu}_l})&=&\frac{G_F}{\sqrt{2}}V_{ub}\bar{ u}\gamma_{\alpha}(1-\gamma_5)b \bar l\gamma^{\alpha}(1-\gamma_5)\nu_l \, ,\nonumber \\
 {\cal H}_{eff}(b\to s l {\bar{l}} )&=&{\frac{G_{F}}{\sqrt{2}}}
V_{tb}V_{ts}^{*}\left[C_{9}^{eff}\left( \mu\right) \bar{s}\gamma _{\alpha}(1-\gamma _{5})b\,\bar{l}\gamma ^{\alpha }(1-\gamma_{5})l+C_{10}(\mu)\bar{s}\gamma _{\alpha}
(1-\gamma _{5})b \bar{l}\gamma ^{\alpha }\gamma _{5} l \right. \nonumber \\
&&\left.-\frac{2m_{b}C_{7}\left( \mu\right) }{q^{2}}\sigma _{\alpha \beta}(1-\gamma _{5})q^{\beta }b \bar{l}\gamma ^{\alpha }l \right]  \, ,
 \end{eqnarray}
where the $G_F$ is the
Fermi constant, the $V_{ub}$, $V_{tb}$, $V_{ts}$ are the CKM matrix elements, $l=\left( e,\mu ,\tau \right) $,
$C_9^{eff}(\mu) = C_9(\mu)+Y_{SD}+Y_{LD}$ , $\mu=m_b$ \cite{BurasRMP}, the $Y_{SD}$ describes the short-distance
contributions from the four-quark operators far away from the $\bar{c}c$
resonances \cite{BurasRMP}, the $Y_{LD}$ denotes the long-distance contributions  from
the four-quark operators near the $\bar{c}c$ resonances and is neglected  due to the absence of
experimental data on the decays $B \to J/\psi S$.
The non-factorizable effects from the $c$-quark loop can
 be absorbed into the effective Wilson
coefficient $C_7^{eff}$  \cite{Gengda}, $C_7^{eff}(\mu) = C_7(\mu)+C'_{b\to s\gamma}(\mu)$,
the $C'_{b\rightarrow s\gamma}$ denotes the absorptive part for the $b \to s
c \bar{c} \to s \gamma$ re-scatterings.

From the effective Hamiltonians, we obtain the partial  decay widths $d\Gamma/dq^2 $, the numerical values  are shown explicitly in Fig.2,
then carry out the integrals
$\tau_B\int dq^2 d\Gamma/dq^2 $ to obtain the branching ratios, where the $\tau_B$ is the lifetime of the $B$-meson,
\begin{eqnarray}
{\rm{Br}}\left(\bar{B}^0\to a_0^+(1450)e \bar{\nu}_e \right)&=&1.59\times 10^{-4} \, , \nonumber\\
{\rm{Br}}\left(\bar{B}^0\to a_0^+(1450)\mu \bar{\nu}_{\mu} \right)&=&1.59\times 10^{-4}\, , \nonumber\\
{\rm{Br}}\left(\bar{B}^0\to a_0^+(1450)\tau \bar{\nu}_{\tau} \right)&=&5.83\times 10^{-5}\, , \nonumber\\
{\rm{Br}}\left(B^-\to K_0^{*-}(1430)e^+ e^- \right)&=&4.51\times 10^{-7}\, , \nonumber\\
{\rm{Br}}\left(B^-\to K_0^{*-}(1430)\mu^+ \mu^- \right)&=&4.48\times 10^{-7}\, , \nonumber\\
{\rm{Br}}\left(B^-\to K_0^{*-}(1430)\tau^+ \tau^- \right)&=&7.35\times 10^{-9}\, .
\end{eqnarray}
In Table 3, we also present the branching ratios from
 the QCD sum rules \cite{Aliev0710},  the perturbative QCD \cite{Lu0811}, the light-cone QCD sum rules
\cite{F-WangYM},   and the covariant light-front quark model \cite{F-Chen},
the values differ widely. The present  predictions of the partial decay widths and the branching ratios can be confronted with the
experimental data at the LHCb in the future.  Furthermore, we expect to extract
 those form-factors from  the semi-leptonic decays $B\to  Sl\bar{\nu}_{l}$, $S\bar{l}l$, and obtain severe constraints on
the input parameter $\lambda_B$ of the $B$-meson light-cone distribution amplitudes.

\section{Conclusion}
In this article, we take the scalar mesons $K^*_0(1430)$ and
$a_0(1450)$ as the conventional two-quark states,  and calculate the $B \to K^*_0(1430)\,
, \, a_0(1450)$ form-factors  $A_1(q^2)$, $A_0(q^2)$,  $T(q^2)$  and
$T_5(q^2)$  with the $B$-meson light-cone QCD sum rules,
   then take those form-factors as the  basic input  parameters to study
the semi-leptonic decays $\bar{B}^0\to a_0^+(1450)l \bar{\nu}_l $  and
 $B^-\to K_0^{*-}(1430)l\bar{l}$, the predictions can be confronted with  the experimental data at the LHCb in the future,
 which are worthy  in identifying the nature  of the scalar mesons.
  In calculations, we observe that the main uncertainty comes from the parameter $\omega_0$ (or $\lambda_B$),
which determines  the line shapes of  the two-particle and three-particle $B$-meson
light-cone distribution amplitudes. We can extract
 those form-factors from the experimental data on the
semi-leptonic decays $B\to  Sl\bar{\nu}_{l}$, $S\bar{l}l$, and
obtain severe constraints on  the input parameter $\lambda_B$.

\section*{Acknowledgments}
This  work is supported by National Natural Science Foundation of
China, Grant Number 11075053,  and the
Fundamental Research Funds for the Central Universities.

\end{document}